\documentclass[12pt]{article}

\usepackage{graphicx}
\usepackage{a4}

\usepackage{fancybox}
\usepackage{epsfig}
\usepackage{color}

\usepackage{amsfonts}
\usepackage{mathrsfs}
\usepackage{amssymb}
\usepackage{amsmath}

\usepackage{dcolumn}
\usepackage{bm}

\def\pa{\partial}

\def\al{\alpha}
\def\be{\beta}

\def\de{\delta}
\def\ep{\epsilon}

\def\la{\lambda}

\def\si{\sigma}

\def\Si{\Sigma}

\def\Om{\Omega}

\newcommand{\ben}{\begin{equation}}
\newcommand{\een}{\end{equation}}
\newcommand{\bea}{\begin{eqnarray}}
\newcommand{\eea}{\end{eqnarray}}
\newcommand{\ba}{\begin{array}}
\newcommand{\ea}{\end{array}}
\newcommand{\bit}{\begin{itemize}}
\newcommand{\eit}{\end{itemize}}

\newcommand{\vs}[1]{\vspace{#1 mm}}

\newcommand{\dsl}{\pa \kern-0.5em /}

\begin{document}

\topmargin 0pt \oddsidemargin 0mm

\begin{flushright}

USTC-ICTS-11-14 \\

\end{flushright}

\vspace{2mm}

\begin{center}

{\Large \bf Strings from geometric tachyon in Rindler space and
black hole thermodynamics}

\vs{10}

 {\large Huiquan Li \footnote{E-mail: hqli@ustc.edu.cn}}

\vspace{6mm}

{\em

Interdisciplinary Center for Theoretical Study\\

University of Science and Technology of China, Hefei, Anhui 230026, China\\

}

\end{center}

\vs{9}

\begin{abstract}
The dynamics of a probe particle or wrapped brane moving in the
two-dimensional Rindler space can be described by a time-dependent
tachyon field theory. Using knowledge of tachyon condensation, we
learn that the infalling brane gets thermalised and produces open
string pairs at the Hagedorn temperature when entering into the
near-horizon Rindler wedge. It is shown that the Hagedorn
temperature of the infalling brane is equal to the Hawking
temperature of the host black hole detected in the same time
coordinate. The infalling brane will decay completely into closed
strings, mainly massive modes, when it reaches the horizon in
infinitely long time as observed by observers at spatial infinity.
Preliminary estimates indicate that the degeneracy of states of the
closed strings emitted from the infalling brane should be
responsible for the increased entropy in the host black hole due to
absorption of the brane.
\end{abstract}



\section{Introduction}
\label{sec:introduction}

String theory as a quantum gravity has been proven successful in
providing microscopic interpretation to black hole thermodynamics.
The Bekeinstein-Hawking entropy is reproduced in some extremal
D-branes by counting the underlying string states
\cite{Strominger:1996sh}. The perfect agreements are further found
by extending to the case of near-extremal black branes with small
temperature \cite{Callan:1996dv,Horowitz:1996fn}, which can be taken
as a linear excitation above the ground state of extremal branes.
The Hawking radiation accounting for the temperature is interpreted
as open strings colliding to forming massless closed strings
\cite{Callan:1996dv}.

However, it is hard to go beyond to study the generical non-extremal
black holes or branes away from extremality. For these black holes,
the near-horizon geometry is the Rindler space. The linear
perturbative analysis does not apply because of strong
backreactions. In this case, the relevance of the thermodynamics to
closed string tachyon has been realised in
\cite{Dabholkar:2001if,Horowitz:2005vp}.

In this work, we try to explore the stringy understanding of the
thermodynamics of non-extremal black holes or branes by dropping in
probe branes, since a large black hole can be formed by swallowing
material repeatedly starting from a small one. The study of the
infalling process of material is directly relevant to the
information loss problem \cite{Hawking:1976ra}. Moreover, probe
matter branes are also widely adopted in holographic models of QCD,
condensed matter and other topics at finite temperature.

We show that a probe particle or wrapped brane moving in the
near-horizon Rindler space can be described by the time-dependent
tachyon field theory. So, from the effective theory or the boundary
conformal field theory (BCFT)
\cite{Sen:2002nu,Sen:2002in,Sen:2004nf} of rolling tachyon, we learn
that the probe brane will experience a Hagedorn transition when it
falls into the Rindler wedge. Open strings
\cite{Strominger:2002pc,Gutperle:2003xf,Maloney:2003ck} and closed
strings \cite{Okuda:2002yd,Lambert:2003zr} will be produced from the
infalling brane at the Hagedorn temperature, which is equal to the
Hawking temperature of the host black hole. The infalling brane will
decay completely into closed strings as the horizon is reached. The
degeneracy of states of the emitted closed strings contributes the
same order of entropy increased in the host black hole due to
absorption of the brane.

We first present the tachyon field configurations that describe the
dynamics of a probe moving in near-horizon Rindler space in
Section.\ \ref{sec:tac-rin}. In Section.\ \ref{sec:stringcreation},
we introduce the results on string production from the rolling
tachyon obtained in the corresponding BCFT and explore their
implications to black hole thermodynamics. Conclusions are made in
the last section. In this paper, we adopt the unit $\al'=1$.

\section{Geometrical tachyon in Rindler space}
\label{sec:tac-rin}

For non-extremal black holes or black $p$-branes, the geometry near
the event horizon is generically Rindler space in the time and
radial directions. In this section, we take two examples to
demonstrate that a probe particle or wrapped $q$-brane moving in
this spacetime can be described by a time-dipendent tachyon field
theory.

\subsection{Schwarzschild black hole}\label{subsec:sch}

We start with the simple and well studied case, the Schwarzschild
black hole:
\begin{equation}
 ds^2=-\left(1-\frac{r_0}{r}\right)dt^2+\left(1-\frac{r_0}{r}
\right)^{-1}dr^2+r^2d\Om_2^2,
\end{equation}
where $r_0=2M$ and $M$ is the mass of the black hole. The Hawking
temperature and the entropy are respectively
\begin{equation}\label{e:tement}
 T_H=\frac{1}{8\pi M}, \textrm{ }\textrm{ }\textrm{ }
S_{BH}=\frac{A}{4}=4\pi M^2.
\end{equation}
The Hawking radiation can be derived by introducing a scalar $\phi$
with mass $m$ (a simple situation is $m=0$) in the black hole
background. Defining the tortoise coordinate $r_*=r+r_0\ln(r/r_0-1)$
and separating out the angular components, one obtains the following
Schr\"{o}dinger equation:
\begin{equation}\label{e:schshreq}
 [\pa_t^2-\pa_{r_*}^2+W(r_*)]\phi(r_*,t)=0,
\end{equation}
where $W(r_*)=(1-r_0/r)[r_0/r^3+l(l+1)/r^2+m^2]$. Approaching the
horizon $r\rightarrow r_0$, $r_*\rightarrow-\infty$ and $W\sim
e^{r_*/r_0}$. Thus, the metric near horizon is approximately:
$ds_2^2\simeq e^{r_*/r_0}(-dt^2+dr_*^2)$. For $r\rightarrow\infty$,
$r_*\rightarrow\infty$ and $W\rightarrow l(l+1)/r_*^2$ in the
massless case.

The near horizon geometry along the time and radial directions is
the Rindler space: $ds_2^2=-\rho^2dt^2+4r_0^2d\rho^2$, with the
coordinate redefinition $r=r_0(1+\rho^2)$ ($\rho$ is dimensionless).
The Hawking temperature $T_H$ given in Eq.\ (\ref{e:tement}) arises
from the periodicity of the Euclidian time coordinate:
$t/(2r_0)\rightarrow t/(2r_0)+2\pi i$. The dynamics of a test
particle moving towards the horizon along the radial direction is
governed by
\begin{equation}\label{e:bosaction}
 S_0=-m_0\int dt\sqrt{\rho^2-4r_0^2(\pa_t\rho)^2}
=-\int d x^0 V(T)\sqrt{1-\dot{T}^2},
\end{equation}
where
\begin{equation}\label{e:boscorpot}
 x^0=\frac{t}{r_0}, \textrm{ }\textrm{ }\textrm{ }
T=-2\ln\rho, \textrm{ }\textrm{ }\textrm{ } V(T)=r_0m_0e^{-T/2}.
\end{equation}
We denote $\dot{T}=\pa_{x^0}T$. In the $x^0$ coordinate, the Hawking
temperature becomes $1/(4\pi)$ due to $x^0/2\rightarrow x^0/2+2\pi
i$. Approaching the horizon $\rho\rightarrow0$, we have
$T\rightarrow\infty$ and $V(T)\rightarrow0$. This action is the
tachyon field action with the potential: $V(T)=T_0/\cosh{(T/2)}$
($T_0=m_0r_0/2$) at late times when the tachyon field $T$ is large
\cite{Sen:2004nf}. The latter is derived in open string field theory
\cite{Kutasov:2003er}. It describes the decaying process of an
unstable D-particle in bosonic string theory. Such a construction of
tachyon field theory is similar to the spirit of the geometrical
tachyon configuration near NS5-branes
\cite{Kutasov:2004dj,Kutasov:2004ct}. The tachyonic instability of
D0-brane probing Shcwarzschild horizon was also noticed in
\cite{Kabat:1998vc}.

If there is no energy loss from the rolling tachyon, the energy
density should be constant
\begin{equation}
 T_{00}=\frac{V(T)}{\sqrt{1-\dot{T}^2}}=C,
\end{equation}
which gives $\ddot{T}=V^2/(2C^2)$. Since we are focusing on the
dynamics near horizon and do not know the information before it
entering into the Rindler space, we can take the constant $C$ as the
energy density when the particle enters the Rindler space. In the
near-horizon region, the infalling particle behaves like an unstable
particle at late times with large $T$. So the solution is
approximately: $T\simeq x^0+ae^{-x^0}+\cdots$, with
$a=(r_0m_0/(\sqrt{2}C))^2$. The pressure is
\begin{equation}
 T_{ij}
=-\frac{V^2}{C}\de_{ij} \simeq-\frac{r_0^2m_0^2}{C}
e^{-x^0}\de_{ij}.
\end{equation}
Thus, the tachyon condensation leads to a pressureless state
approaching the horizon, which corresponds to non-relativistic,
massive closed strings \cite{Sen:2003bc,Sen:2004nf}. This will be
addressed in the next section.

\subsection{Black branes}\label{subsec:brane}

We next consider the black brane gravitational solutions
\cite{Horowitz:1991cd,Duff:1993ye} in type II string theories.
Redefine the coordinate $r^{7-p}=(1+\rho^2)r_+^{7-p}$ so that the
event horizon is located at $\rho=0$. In non-extremal case, the
black $p$-brane ($p<7$) metric near the event horizon 
takes the form
\begin{equation}\label{e:branesol}
 ds^2=\ep^{-\frac{7-p}{2}}\left[-\rho^2dt^2+\ep^{7-p}d\vec{x}^2
+\frac{4r_+^2}{(7-p)^2}\ep^{p-5}d\rho^2+r_+^2\ep^2d\Om^2_{8-p}
\right],
\end{equation}
\begin{equation}
 e^\Phi=g_s\ep^{\frac{(p-3)(7-p)}{4}}.
\end{equation}
where the parameter $\ep$ is defined as the discrepancy between the
radii $r_\pm$ of the inner and outer horizons:
$\ep^{7-p}=1-(r_-/r_+)^{7-p}$. The radii $r_\pm$ are related to the
tension and the RR charge of black branes. The extremal limit is
$\ep\rightarrow0$ with $r_-=r_+$, in which the tension and charge
are equal. When the black brane is chargeless, $r_-=0$ and so
$\ep=1$.

A D-particle or D$q$-brane wrapped on circles moving along the
radial direction is described by the action:
\begin{equation}\label{e:supaction}
 S_q=-\tau_q\int dt e^{-\Phi}\sqrt{-\det{G}}=\int dx^0
V(T)\sqrt{1-\dot{T}^2},
\end{equation}
where
\begin{equation}\label{e:supcor}
 T=-\sqrt{2}\ln\rho, \textrm{ }\textrm{ }\textrm{ }
x^0=\frac{7-p}{\sqrt{2}r_+}\ep^{\frac{5-p}{2}}t,
\end{equation}
\begin{equation}\label{e:suppot}
 V(T)=2T_q e^{-T/\sqrt{2}},
\textrm{ }\textrm{ }\textrm{ }
T_q=\frac{r_+}{\sqrt{2}(7-p)}g_s^{-1}\tau_q\ep^{\frac{p^2-7p+4}{4}}.
\end{equation}
The Hawking temperature of the host black brane (\ref{e:branesol})
detected in the $x^0$ coordinate is $1/(2\sqrt{2}\pi)$ due to the
periodicity $x^0/\sqrt{2}\rightarrow x^0/\sqrt{2}+2\pi i$ in the
Rindler space: $ds_2^2=(\textrm{const.})[-\rho^2(dx^0)^2+2d\rho^2]$.
The tachyon potential $V(T)$ is the large $T$ case of the potential
in superstring theory: $V(T)=T_q/\cosh(T/\sqrt{2})$ obtained in
\cite{Kutasov:2003er}. It is constructed like this because we are
discussing the dynamics in the black brane background
(\ref{e:branesol}) obtained in type II superstring theories.

In this action, the non-vanishing energy-momentum tensor at late
times evolves as
\begin{equation}\label{e:suptensor}
 T_{00}=C, \textrm{ }\textrm{ }\textrm{ }
T_{ij}\simeq\frac{4T_q^2}{C}e^{-\sqrt{2}x^0}\de_{ij},
\end{equation}
with the tachyon solution: $T\simeq x^0+ae^{-\sqrt{2}x^0}+\cdots$,
where $a=\sqrt{2}(T_q/C)^2$. We also get pressureless matter towards
the end of condensation as $x^0\rightarrow\infty$.

\section{Strings from tachyon and black hole thermodynamics}
\label{sec:stringcreation}

It is known that the tachyon effective action with the potential
$V(T)=T_q/\cosh(\be T)$ ($\be=1/2$ for bosonic string and
$=1/\sqrt{2}$ for superstring) has a corresponding $c=1$ BCFT
description \cite{Okuyama:2003wm,Lambert:2003zr,Sen:2004nf}. The
BCFT worldsheet action is \cite{Sen:2002nu,Sen:2002in}
\begin{equation}\label{e:BCFTaction}
 S=-\frac{1}{2\pi}\int_{\Si} d^2z\pa X^\mu\bar{\pa}
X_\mu+\int_{\pa\Si}d\tau m^2(X^0(\tau)),
\end{equation}
where $\tau$ labels the coordinate on the boundary of the
worldsheet. In bosonic string theory, the boundary interaction term
$m^2(X^0)$ is $(\la/2)e^{X^0}$ (for $C=T_q$) or $\la\cosh X^0$ (for
general $C$). The latter gives rise to a full S-brane, which
describes the creation and then decaying processes of an unstable
D-brane, while the former gives rise to a half S-brane, which
describes the decaying process of an unstable D-brane. In
superstring theory, the tachyon profile takes the form:
$\la\cosh(X^0/\sqrt{2})$.

In order to understand the near-horizon behaviour of a probe brane
discussed in the previous section, we only need to concentrate on
the late-time evolution of the rolling tachyon. The non-vanishing
components of the energy-momentum tensor at late times obtained from
evaluation on the $X^0$ component of the boundary states are given
by (in the $g_s\rightarrow0$ limit) \cite{Sen:2002nu,Sen:2002in}:
\begin{equation}\label{e:bostensor}
T_{00}=\frac{1}{2}T_q[1+\cos(2\la\pi)], \textrm{ }\textrm{ }\textrm{
}
 T_{ij}\simeq
 \left\{
 \begin{array}{cl}
 -\frac{1}{\sin(\la\pi)}T_qe^{-x^0}\de_{ij},  \textrm{ }\textrm{ }\textrm{
} (\textrm{bosonic string}) \\
 -\frac{1}{\sin^2(\la\pi)}T_qe^{-\sqrt{2}x^0}\de_{ij}, \textrm{ }\textrm{ }
\textrm{ } (\textrm{superstring})
 \end{array}
 \right.
\end{equation}
where $T_q$ is the tension of the unstable D$q$-brane and $0\leq
i,j\leq q$. Hence, the tachyon BCFT (\ref{e:BCFTaction}) is
comparable to the tachyon effective actions constructed in the time
coordinate $x^0$ in the previous section. The string production from
decaying tachyon based on the BCFT has been studied previously. In
what follows, we investigate the implications of the results to the
thermodynamics of non-extremal black holes.

\subsection{Open string pair creation: Hagedorn and Hawking temperatures}

The creation of open string pairs from the rolling tachyon with
non-vanishing $g_s$ is discussed in
\cite{Strominger:2002pc,Gutperle:2003xf,Maloney:2003ck} in the
context of S-branes. In their treatments, a minisuperspace
approximation is adopted, i.e., only zero modes
$x^\mu=(\hat{t},\vec{x})$ of $X^\mu=(X^0,\vec{X})$ are considered.
If assuming $x^\mu$ to be independent of the worldsheet coordinate
$\si$, we can obtain an effective action from the action
(\ref{e:BCFTaction}), from which the Hamiltonian can be derived. We
thus can get the Klein-Gordon equation for the wave function
$\psi(t,\vec{x})$ of free open strings by making the constraint
$L_0+\widetilde{L}_0=0$:
\begin{equation}
 [\pa_{\hat{t}}^2-\vec{\nabla}_{\vec{x}}^2+2m^2(\hat{t})+(N-1)]
\psi(\hat{t},\vec{x})=0,
\end{equation}
where $N$ is the contribution from oscillators. This equation with
exponentially increasing $m^2(\hat{t})$ is similar to Eq.\
(\ref{e:schshreq}) in the near-horizon region with $W\sim
e^{r_*/r_0}$. The wave function can be expanded in the way:
$\psi(\hat{t},\vec{x})=u(\hat{t})e^{i\vec{p}\cdot\vec{x}}$. The
calculation of the in and out states indicates that open string
pairs are created at $\hat{t}\rightarrow\infty$ stating from a
vacuum with no incoming open strings at $\hat{t}\rightarrow-\infty$.
The energy spectrum of the created open strings is thermal with the
Hagedorn temperature, which we denote as $T_{Hag}^{(x^0)}$, since
this temperature is derived in the BCFT or the field theories
(\ref{e:bosaction}) and (\ref{e:supaction}) constructed in the $x^0$
coordinate. For observers at spatial infinity, the Hagedorn
temperature should be the one observed in the black hole coordinate
$t$. We denote the temperature in this frame as:
$T_{Hag}^{(t)}=T_{Hag}^{(x^0)}(\pa x^0/\pa t)$.

For bosonic string case, the open strings are created at the
Hagedorn temperature $T_{Hag}^{(x^0)}=1/(4\pi)$. As stated in
Section.\ \ref{subsec:sch}, this temperature is also the Hawking
temperature detected in the $x^0$ coordinate. Therefore, the
Hagedorn temperature for open string creation on the infalling
particle in the $t$ coordinate is exactly the Hawking temperature
$T_H$ (also detected in $t$) of the host black hole. In summary, the
relation between the two temperatures can be expressed as follows
\begin{equation}\label{e:hawhagbos}
 T_H=T_{Hag}^{(t)}=\frac{1}{r_0}T_{Hag}^{(x^0)},
\textrm{ }\textrm{ }\textrm{ }\textrm{ }
T_{Hag}^{(x^0)}=\frac{1}{4\pi},
\end{equation}
where $T_H$ is given in Eq.\ (\ref{e:tement}) for Schwarzschild
black hole. This means that the infalling material gets thermalised
at the Hagedorn temperature or the Hawking temperature when it
enters into the near-horizon Rindler space.

As noted in \cite{Maloney:2003ck}, this Hagedorn temperature for
open string creation essentially arises from the periodicity of
$X^0$ (or $\hat{t}$) in the boundary term $\la e^{\pm X^0}$:
$X^0\rightarrow X^0+2\pi i$. More generally, we have mixed thermal
states due to $X^0\rightarrow X^0+2n\pi i$, with positive integer
$n$. Thus, the Hagedorn temperature $T_{Hag}^{(x^0)}=1/(4\pi)$ is
just the case of $n=2$ for bosonic string theory.  Similar mixed
states for the host black hole can also be constructed from the
periodicity of the imaginary time $x^0$.

For superstring theory, the obtained Hagedorn temperature is:
$T_{Hag}^{(x^0)}=1/(2\sqrt{2}\pi)$, which arises from the
periodicity: $X^0\rightarrow X^0+2\sqrt{2}\pi i$ in the boundary
term $\la e^{\pm X^0/\sqrt{2}}$. The temperatures for mixed states
are: $T_{Hag}^{(x^0)}=1/(2\sqrt{2}n\pi)$ ($n\geq1$). On the other
hand, the Hawking temperature detected in the $x^0$ coordinate is
also $1/(2\sqrt{2}\pi)$ from $x^0\rightarrow x^0+2\sqrt{2}\pi i$, as
have been discussed in Section \ref{subsec:brane}. So, once again,
the Hagedorn temperature for open string creation on the probe brane
is equal to the Hawking temperature of the host black brane
(\ref{e:branesol}):
\begin{equation}
 T_H=T_{Hag}^{(t)}=\frac{7-p}{\sqrt{2}r_+}\ep^{\frac{5-p}{2}}
T_{Hag}^{(x^0)}, \textrm{ }\textrm{ }\textrm{ }\textrm{ }
T_{Hag}^{(x^0)}=\frac{1}{2\sqrt{2}\pi}.
\end{equation}

It is worth pointing out that the equivalence between the two
temperatures is independent of the specific construction of tachyon
effective actions. All the temperatures actually arise from the
unique periodicity of the Euclidian time coordinate in Rindler
space. If we recover the factor $1/\sqrt{\al'}$ before $X^0$ in the
boundary term, then there should be a corresponding $1/\sqrt{\al'}$
before $T$ in $V(T)$ in the effective theory (in this case, both
$X^0$ and $T$ become of dimension length). Accordingly, the period
of imaginary $x^0$ should be multiplied by a factor $\sqrt{\al'}$.
The resulting temperatures are still the same.

\subsection{Closed string emission: the increased entropy}

The Hagedorn temperature is the maximum temperature of a system and
it usually signals phase transition. The energy of the rolling
tachyon should be converted into closed string radiation due to the
exponentially increasing open string mass. Since the Hagedorn
temperature for open string pair creation on the unstable $q$-brane
is equal to the Hawking temperature of the host black hole, the
absorption and the emission of massless closed strings on the
$q$-brane should be in equilibrium, with the same rate. So the
emission of massless closed strings from the rolling tachyon
discussed in \cite{Chen:2002fp,Rey:2003xs} should be suppressed.
Actually, it has been found in \cite{Okuda:2002yd,Lambert:2003zr}
that low dimensional unstable D-branes indeed decay into mainly very
massive closed string states. On the other hand, closed strings can
also split into open strings at the Hagedorn temperature
\cite{BatoniAbdalla:2007zv}.

In the leading order of $g_s$, the total average energy and number
(per spatial worldvolume $V_q$) of emitted closed strings with large
level number $n$ are calculated to be \cite{Lambert:2003zr}:
\begin{equation}
 \frac{\bar{E}}{V_q}=(2\pi)^{-q}\int dEE^{-\frac{q}{2}},
\textrm{ }\textrm{ }\textrm{ }
\frac{\bar{N}}{V_q}=(2\pi)^{-q}\int\frac{dn}{2n}(4n)
^{-\frac{q}{4}},
\end{equation}
in either bosonic string or superstring theory. For an unstable
D$q$-brane ($q\leq2$), the emitted energy diverges. We can choose
the energy cut-off at the order of the mass or tension of the
unstable particle or D$q$-brane. This means that the energy
completely converts to the closed strings, with most energy
transferred into the very massive states, towards the end of
condensation. In the present context, this implies that the
infalling wrapped $q$-brane ($q\leq2$) will dissolve into closed
strings when reaching the horizon in infinitely long time as
observed at infinity.

Let us first consider dropping a particle into the Schwarzschild
black hole. On the macroscopic side in classical gravity, the black
hole mass $M$ increases by $\de M\sim m_0$. When entering into the
near-horizon region, this particle behaves like a decaying unstable
particle. On the microscopic side in the tachyon field theory or
BCFT, the particle completely decays into closed strings, mainly
massive modes, when it approaches the horizon. In terms of the first
law, the black hole entropy $S$ should increase by $\de S\sim\de
M/T_H$ due to capture of the particle (ignoring the charge and
angular momentum contributions). This increased entropy should be
accounted for by information stored in the closed strings that the
particle decays into, as will be examined below.

Involving the relation (\ref{e:hawhagbos}), the first law then
reads:
\begin{equation}\label{e:bosfirlaw}
 \de S\sim \frac{m_0}{T_{H}}=\frac{r_0m_0}{T_{Hag}^{(x^0)}}.
\end{equation}
This is the Hagedorn growth rule of entropy relying on mass commonly
seen in string theory \cite{Green:1987sp}. This equation says that
the increased entropy should be the ratio of the corresponding mass
($m_0$ or $r_0m_0$) of the test particle to the corresponding
Hagedorn temperature ($T_{Hag}^{(t)}=T_H$ or $T_{Hag}^{(x^0)}$)
respectively observed in the $t$ or $x^0$ coordinate.

We now check whether the degenerate states of the emitted closed
strings can account for the increased entropy in the host black hole
required by the first law (\ref{e:bosfirlaw}). The density of states
of the emitted closed strings is obtained in \cite{Lambert:2003zr}
and it takes the form of the one of bosonic open strings in Hilbert
space
\begin{equation}\label{e:dnbosop}
 d_n\sim n^{-\frac{27}{4}}e^{4\pi\sqrt{n}}.
\end{equation}
with $E^2\sim4n$ at large level $n$. The energy $E$ is the order of
the mass $r_0m_0$ of the particle measured in the $x^0$ coordinate.
So the entropy contributed by the closed string states is
\begin{equation}
 \ln d_n\sim \frac{r_0m_0}{2T_{Hag}^{(x^0)}}-\frac{27}{2}
\ln\left(\frac{r_0m_0}{2}\right).
\end{equation}
Thus, the leading term gives half of the result required by the
first law in Eq.\ (\ref{e:bosfirlaw}). The $1/2$ factor may arise
from approximations, like the left-right mover identical assumption,
adopted in \cite{Lambert:2003zr} or in the above treatments. If we
use the expression of closed string degenerate states in stead of
the open string one (\ref{e:dnbosop}):
$d_n^{(\textrm{clsoed})}=(d_n^{(\textrm{open})})^2$, the result will
agree with (\ref{e:bosfirlaw}).

In the black brane backgrounds, the tension of the probe $q$-brane
is dependent on the parameter $\ep$. At infinity, the tension is of
order $\tau_q g_s^{-1}$. Approaching the near horizon region, the
effective tension is $\tau_qg_s^{-1}\ep^{-\frac{(p-2)(7-p)}{4}}$. It
can become very large or small towards the extremal limit, which
means that there can be arbitrarily large or small amount of closed
string radiated at the end of tachyon condensation. But, for a probe
$q$-brane with given tension at infinity, the closed strings emitted
from it should be limited. So there should be large deviation in the
case $\ep\rightarrow0$, in which there should be a different or
revised theory to describe the dynamics. Therefore, we only consider
the case away from extremality with $\ep\sim\mathcal{O}(1)$. In this
case, the increased entropy in the black brane due to absorption of
a wrapped $q$-brane can be expressed as, following the first law
\begin{equation}\label{e:firlawsup}
 \de S\sim \frac{\tau_qg_s^{-1}}{T_{H}}
\sim\frac{2T_q}{T_{Hag}^{(x^0)}}.
\end{equation}
where the tension $T_q\sim r_+g_s^{-1}\tau_q$ of the unstable
$q$-brane is defined in Eq.\ (\ref{e:suppot}). In superstring
theory, the density of states should grow slower. If the degenerate
state is still the open string one: $d_n\sim
n^{-11/4}\exp(2\pi\sqrt{2n})$ \cite{Green:1987sp}, we shall also get
half of the result (\ref{e:firlawsup}).

\section{Conclusions}
\label{sec:conclusion}

The dynamics of infalling material in the near-horizon Rindler space
can be described by the time-dependent tachyon field theory, in a
holographic sense. Hence, the infalling material will experience a
Hagedorn transition when falling through the near-horizon region of
non-extremal black holes or branes. It will unavoidably dissolve
into closed strings as approaching the horizon. This transition
should be related to the confinement-deconfinement phase transition
in holographic QCD models.

Our results have two important implications. One is that we should
not automatically assume that any matter can move into the
near-horizon region without any change in itself. As we show, the
material will get thermalised and will decay into something else
(closed strings) in this region. In plenty of previous classical
discussions, this assumption has actually been adopted. Another
implication of our results is that the information should be not
lost when matter collapse into a black hole. The information is
stored in the closed strings emitted from the particle. But here we
do not transplant our discussion into the traditional semi-classical
approach and find out the origin of information loss in there, which
is left for further study.

\section*{Acknowledgements\markboth{Acknowledgements}{Acknowledgements}}

We would like to thank Jianxin Lu for useful comments.

\newpage
\bibliographystyle{JHEP}
\bibliography{b}

\end{document}